\documentclass[twocolumn]{article}
\usepackage{graphicx}
\usepackage{amsmath}
\usepackage{hyperref}
\usepackage{authblk}
\usepackage{placeins}

\title{Sensing Social Media Signals for Cryptocurrency News}
\author[1]{Johannes Beck \footnote{Contributed equally to the first authorship}}
\author[1]{Roberta Huang *}
\author[1]{David Lindner *}
\author[2]{Tian Guo}
\author[3]{Ce Zhang}
\author[2]{Dirk Helbing}
\author[2]{Nino Antulov-Fantulin\thanks{anino@ethz.ch}}
\affil[1]{Department of Computer Science, ETH Zurich}
\affil[2]{Computational Social Science, ETH Zurich}
\date{}

\begin{document}

\maketitle

%
% The abstract is a short summary of the work to be presented in the article.
\begin{abstract}
The ability to track and monitor relevant and important news in real-time is of crucial interest in multiple industrial sectors. In this work, we focus on the set of cryptocurrency news, which recently became of emerging interest to the general and financial audience. 
In order to track relevant news in real-time, we (i) match news from the web with tweets from social media, (ii) track their intraday tweet activity and (iii) explore different machine learning models for predicting the number of article mentions on Twitter within the first 24 hours after its publication. We compare several machine learning models, such as linear extrapolation, linear and random forest autoregressive models, and a sequence-to-sequence neural network. We find that the random forest autoregressive model behaves comparably to more complex models in the majority of tasks.
\end{abstract}

\section{Introduction}
Social media and news play an important role in driving the fluctuation of economic indicators and financial markets~\cite{chakraborty2016predicting}, \cite{hu2018listening}, \cite{Pikorec2014}, \cite{Zhang_tradingstrategies} in a nontrivial fashion. 
Recently, novel financial markets have emerged, that are exchanging from fiat money (USD, EUR, CNY) to cryptocurrencies and vice versa \cite{Guo2018, btc-nmf}. 
As of December 2018, cryptocurrencies have a total market capitalization of \$120 billion, with more than 250000 transactions per day. In 2017 Bitcoin was ranked second on the Google Trends list of popular topics in global news. Despite a decrease of interest towards cryptocurrencies in 2018 according to Google Trends, the number of daily articles related to cryptocurrencies is still notably high.
Therefore, it is not surprising that the rapid development of cryptocurrency has attracted increasing attention from news and social media.

A large volume of news articles about cryptocurrencies, published daily can make it hard for individuals to filter out relevant information and make informed decisions in this domain. Fortunately, people share and discuss news every day in large quantities on social media platforms, e.g. on Twitter, which is the focus of this paper. 
Therefore, social media can be a good proxy to monitor and track ''important'' news about cryptocurrencies.
Our work is motivated by the hypothesis that high engagement with a news article on Twitter is related to the ''importance'' of an article.

%Moreover, news articles and associated discussion on Twitter are extremely time sensitive by nature.
%A on-line data pipeline is highly useful to keep up with the pace of news and its propagation on Twitter. 
%Such a system can open up the door of relating news, twitter and cryptocurrency market data to study their interaction.  

In this paper, we introduce an online data mining system which connects news and tweets discussing it.
We also perform preliminary data exploratory and predictive analytic using machine learning and deep learning. 
Overall, the contribution of this paper is as follows: (i) We build an online data mining pipeline to extract news articles from a discussion on Twitter and collect tweets associated with the articles. This paired news and tweet data is continuously updated in a cloud database.
    This data is a rich source for studying public interest and attention on cryptocurrency and the potential effect of social media on the market.  (ii) Based on the news and associated tweets collected by the pipeline, we perform exploratory data analysis to characterize news discussion on Twitter.
    (iii) We apply machine learning and deep learning models to predict the popularity of news articles on Twitter.
    We aim to predict the number of tweets mentioning an articles related to cryptocurrencies, which we consider as a measure of its ''importance''.

\section{Related work}
% effect on financial networks
%In the first category,
Many studies have focused on the relationship between social media, news, and other information from the www onto financial markets \cite{Chen2012, Andersen2007, Pikorec2014}.
%The effects of different social media forums on Bitcoin returns was studied in \cite{mai2015impacts}. In \cite{kristoufek2013bitcoin}, the authors connected cryptocurrencies with Google Trends and Wikipedia to investigate their relationship. 
%The effect of real-world events obtained from news data to the fluctuations in socio-economic indicators was studied in \cite{chakraborty2016predicting, Pikorec2014}. The authors in \cite{hu2018listening} exploit natural language processing to process news articles for directly predicting the stock trend. 
However, the main focus of our work is modeling and prediction of news popularity via social media. In
\cite{tsagkias2011linking}, the authors link a given news article to social media utterances that implicitly reference it through a dedicated query model. Tracking and automatically connecting news articles to Twitter conversations by Twitter hashtags was studied in \cite{shi2014insight4news}. In \cite{bandari2012pulse}, the authors constructed a multi-dimensional feature space derived from an article and use a conventional SVM to predict its popularity. The authors in \cite{Hawkes} show how the class of temporal point processes (Hawkes) can be used for predicting Retweet dynamics.
The authors in \cite{dou2018predicting} propose how to leverage knowledge base information for improving popularity prediction.
Starting from the idea that only a small amount of news articles become popular, \cite{resampling} focused on the subset of the most popular news to rank articles. In \cite{coldstart} it formulates article importance prediction as a classification task using SVM. %They also define news importance as the number of tweets mentioning an article.

In this paper, we exploit ensemble machine learning and sequence to sequence (seq2seq) deep learning to study the predictability of crypto news popularity on Twitter in real-time mode. In contrast to others, our analysis is focused on the intraday importance prediction.

% As features they use the time of publication, the source of the news, its genre and some linguistic features, such as name entities and a sentiment score. Their best performing classification model is a support vector machine which achieves an accuracy of 79.7\%.

% \cite{retweets} study news propagation on Twitter and develop a prediction model for news popularity on Twitter, while measuring the popularity as the number of retweets. They manage to predict the number of retweets with around $89\%$ correlation to the observed numbers of retweets.

%To the best of our knowledge, news importance prediction has not been investigated as a time-series prediction task on the number of Twitter mentions. We focus on news importance prediction using the observed number of Twitter mentions within 24 hours of an article's publication.

\begin{figure}[ht]
  \centering
  \includegraphics[width=\linewidth]{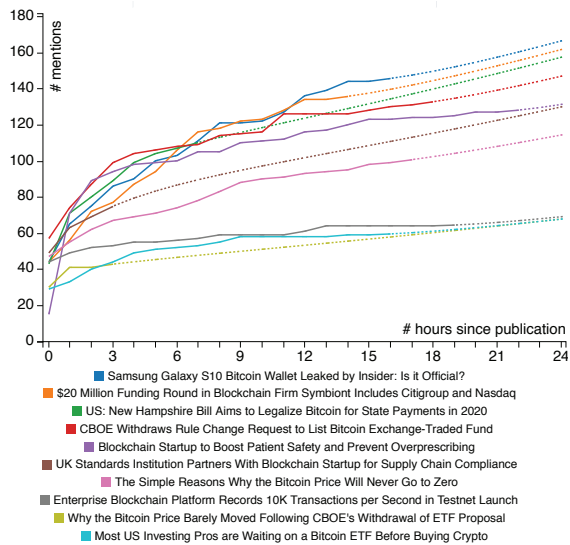}
  \caption{The tracking of crypto news intraday activity on 24th Jan 2018 at 15:28:37 CET, visualized by our server. 
}
 \label{fig:webChart}
\end{figure}

\section{Data pipeline}
%This section describes how we approach the problem of predicting news popularity. 
%In the first subsection we give a more detailed description of the data. Based on this, we formalize the prediction task we aim to solve. Finally, we show how we extract features from the data and specify the models that we use in our experiments. 

%\subsection{Overview of data collection system}\\
%We use two different data sources provided by the COSS group:\footnote{Computational Social Science Group, Department of Humanities, Social and Political Sciences, ETH Zürich} a static dataset of articles and tweets spanning about two weeks, and data from a live database that provides tweets and articles in real-time.

%During the data collection step, performed by the COSS group, 

The data pipeline consists of a real-time online system, with the following components: Twitter collection, article collection, and tweet-article matching. 

\begin{figure}[h]
  \centering
  \includegraphics[width = 0.95\linewidth]{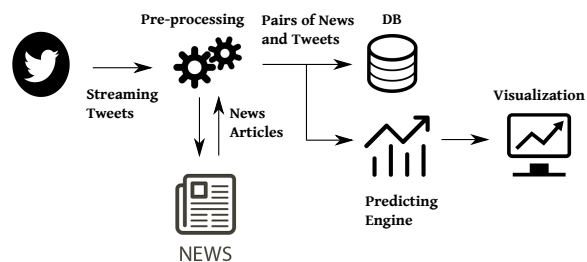}
  \caption{Architecture of the data pipeline.}
\end{figure}

%the tweets are loaded from the publicly accessible Twitter API. The news articles are are filtered with a list of cryptocurrency related keywords. Tweets are then matched to articles if they contain a link towards the article. The Twitter API does only provide a random sample of all tweets at a given time point. Twitter does not publicly disclose how big this sample is.

%During the initial phase of the project the online database was not accessible yet and we worked with the static dataset. Later we obtained access to the live database. As the database was set up after the static dataset was collected, it does not contain the data from the static dataset. To increase the amount of data we can use, we load data from both sources.
%
%\paragraph{Static dataset} The data file is provided in the Python pickle file format\footnote{\url{https://docs.python.org/3/library/pickle.html}} which can easily be loaded using Python standard libraries. The file contains a list of Python dictionaries describing matchings with all the available information about the article and the tweets. Note that we only load the small subset of properties that are relevant for our task (cf. \ref{tab:dat_article,tab:dat_tweet,tab:dat_matching}).
%
% [Tian] The dataset that we used was obtained by crawling hourly feed data from Twitter.com. To ensure that we obtained all tweets referring to a movie, we used keywords present in the movie title as search arguments.

\textbf{The Twitter data collection} was implemented by using the publicly accessible Twitter streaming API with real-time filtering by a list of cryptocurrency related keywords. The Twitter API does only provide a random sample of all tweets.  

\textbf{The article data colletion} is obtained by scraping news from the dynamic set of gazetteer source URLs. The set of gazetteer source URLs is automatically updated by extracting the URLs from the content of downloaded tweets. 

\textbf{The tweet-article matching data} is the document-oriented database online instance, that contains matchings between news articles and tweets. The matching exists if the tweet explicitly contains the URL of an article. 

% too technical -- not important in my opinion
%Each article can have multiple partial matching entries in the database because the data is added in an incremental fashion. Old matching entries are not updated; instead, a new entry is added every time the database is updated. This new entry contains all tweets published since the last update matched to the corresponding articles. As an example, consider an article $a$ and three tweets $t_1, t_2, t_3$ published at timesteps $0$, $1$ and $3$ respectively. Suppose the database receives updates at timestep $2$ and $4$. Then the database contains two partial matching entries for $a$. The first entry relates $a$ to $t_1$ and $t_2$, while the second entry relates $a$ to $t_3$.

%\subsection{Data postprocessing}\label{sec:preprocessing}
Before extracting features from the data, we first process the data for further usage.
In a first step, we merge some of the matchings together. This is done for two reasons. First, the online database only contains incremental matchings which have to be merged together in order to provide a complete matching of the article to tweets. Secondly, the raw URLs of the articles can contain query strings, which often contain information not related to which article the URL refers to. Hence, by changing those parameters one can obtain arbitrarily many different URLs linking to the same article. Because of this, there are often multiple different article entries in the database or data file for the same article. On the other hand, there are also some websites that use the query string to distinguish between articles. Therefore, we merge the matchings of two articles if they fulfill the following 3 conditions:
\begin{enumerate}
    \item The URLs of both articles share the same host as well as the same path.
    \item Both articles have the same title.
    \item Both articles were published at the same time.
\end{enumerate}
These conditions allow for \textbf{merging} of articles of which the URLs have different query strings while the last two conditions prevent the system from merging articles which are distinguished by the query string in the URL. While merging the articles we also remove duplicate entries for the same tweet which are sometimes present in the database.

According to the publication time in the data, some of the articles were published 2000 years ago or even in the future. These publication dates are clearly wrong.
%And even for articles with correct timestamps, it can happen, that the data does not contain sufficient information about the Twitter dynamics of the article. This is the case if the article was published before the first tweet in the dataset or less than 24 hours before the last tweet in the dataset. In those cases, it is impossible to observe the number of mentions within the first 24 hours after the publication of the article. 
We, therefore, \textbf{remove} all articles that were published outside of an acceptable time-interval.\\ %After removing these articles, the data loaded from the file yields a dataset containing tweets published between 2018-10-13 20:44:17 and 2018-10-31 11:37:33, as well as articles published between 2018-10-13 20:50:00 and 2018-10-30 11:36:17. The first available tweet in the online database was published at 2018-11-03 13:35:01 and the first article within the acceptable timerange at 2018-11-03 13:35:14.

\begin{figure}[h]
  \centering
  \includegraphics[width=0.95\linewidth]{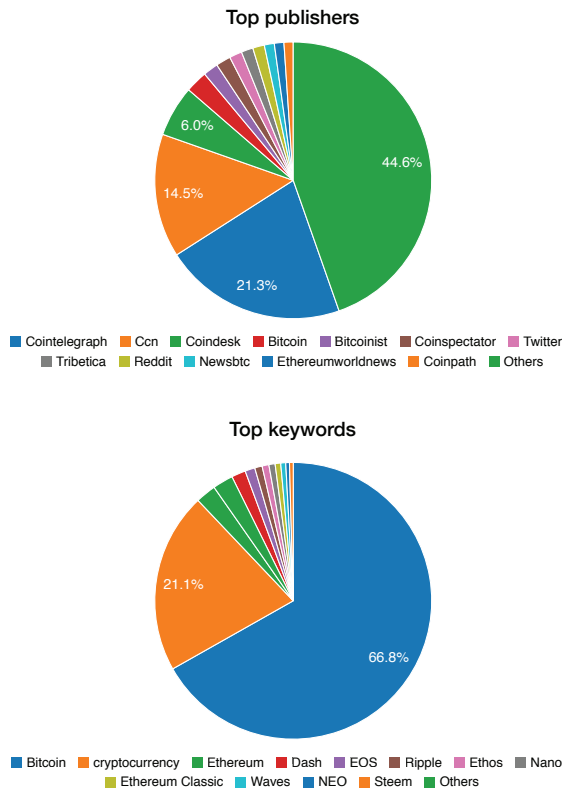}
  \caption{Top publishers and top keywords on 24th Jan 2018 at 15:28:37 CET, visualized by our server. 
}
\label{fig:webStats}
\end{figure}

\begin{figure}[h]
  \centering
  \includegraphics[width=\linewidth]{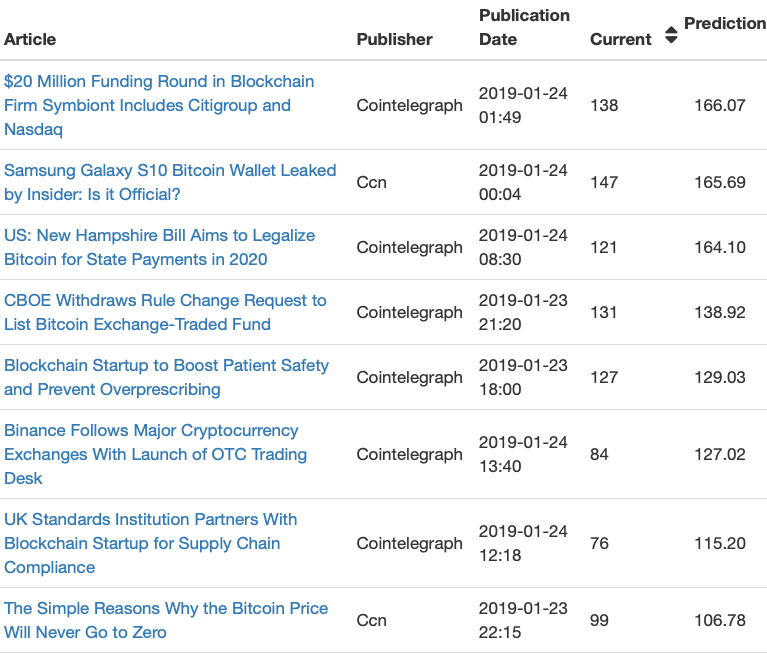}
  \caption{Top article ranked by our system on 24th Jan 2018 at 15:28:37 CET.
}
\label{fig:webTable}
\end{figure}

Let us describe the data, that we are gathering and then go on to describe how we process this data.
We work with three entities in our dataset: news articles, tweets, and matchings, i.e. the relations between articles and tweets. While these concepts are easy to understand intuitively, we specify their attributes here to avoid confusion.

In Tab. \ref{tab:dat_article} and \ref{tab:dat_tweet}, we provide a list of the properties of a news \emph{article} and a \emph{tweet} respectively. A \emph{matching} (see Tab. \ref{tab:dat_matching}) relates to a set of tweets that reference the article.

\begin{table}[htp]
  \caption{Properties of an article entity.}
  \label{tab:dat_article}
  \centering
  \begin{tabular}{ll}
    \texttt{URL}    & The url of the article.    \\
    \texttt{title} & The headline of the article.\\
    \texttt{publication time} & The publication timestamp. \\
    \texttt{text} & The text body of the article. \\
  \end{tabular}
\end{table}

\begin{table}[htp]
  \caption{Properties of a tweet entity.}
  \label{tab:dat_tweet}
  \centering
  \begin{tabular}{ll}
    \texttt{user}    & The Twitter username.    \\
    \texttt{text} & The content of the tweet.\\
    \texttt{publication time} & The timestamp of the tweet. \\
    \texttt{links} & The urls in the text.\\
  \end{tabular}
\end{table}

\begin{table}[htp]
  \caption{Properties of a matching entity.}
  \label{tab:dat_matching}
  \centering
  \begin{tabular}{ll}
    \texttt{article} & An article as described in Tab.\ref{tab:dat_article}.    \\
    \texttt{tweets} & A list of matched tweets. \\
  \end{tabular}
\end{table}

\begin{figure}[ht]
  \centering
  \includegraphics[width=0.95\linewidth]{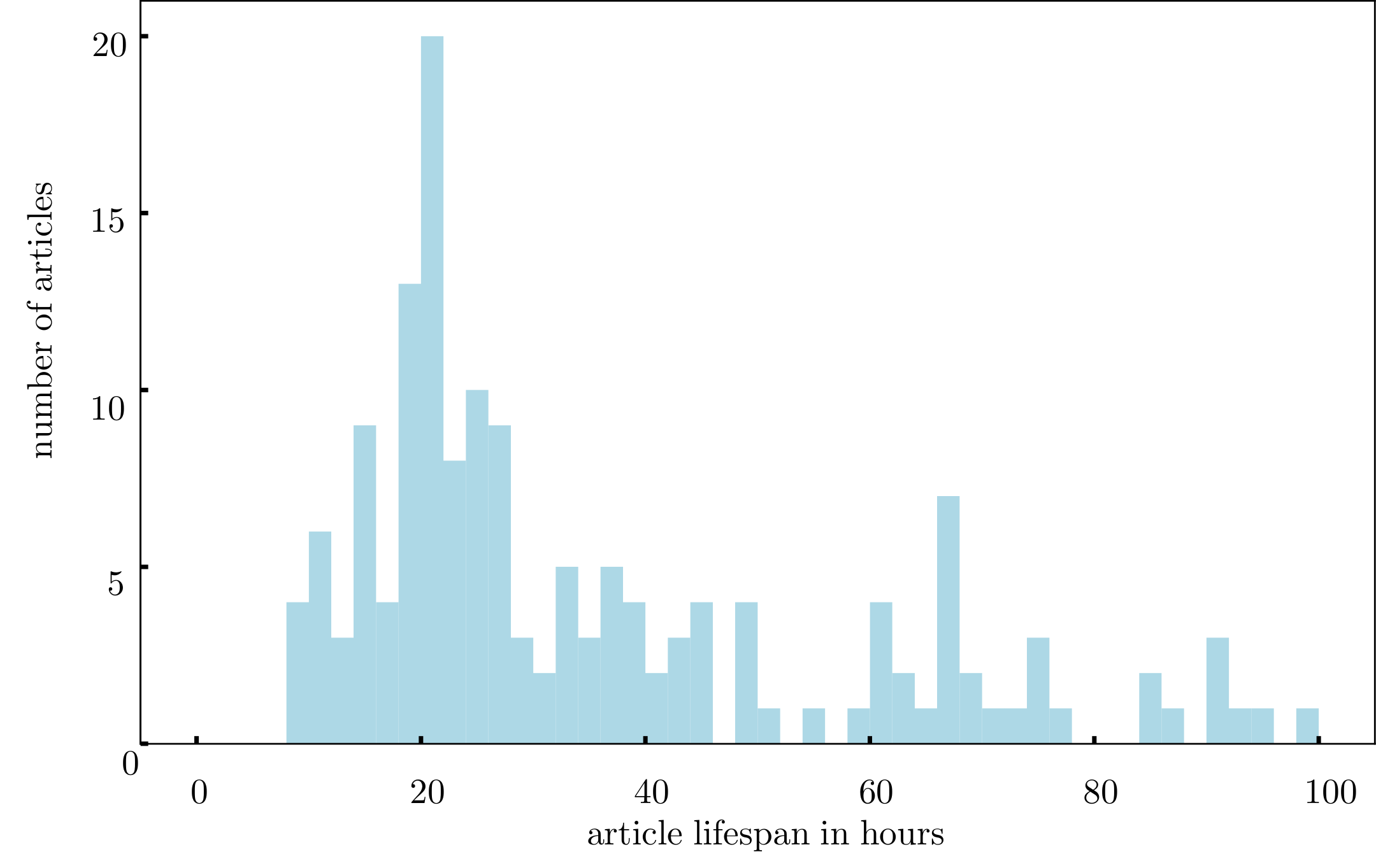}
  \caption{Histogram of article lifespan during one week starting at 13.10.2018, defined as the time after which an article reaches 90\% of its total mentions (on the right). We observe that most articles, with at least 100 mentions, reach the end of their lifespan around 24 hours after publication.
}
 \label{fig:lifespan_hist}
\end{figure}

%\begin{figure}
    %\centering\small
    %\def\svgwidth{\columnwidth}
    %\input{lifespan_hist.pdf}\vspace*{0.5cm}
   % \caption{Histogram of article lifespan during one week starting at 13.10.2018, defined as the time after which an article reaches 90\% of its total mentions (on the right). We observe that most articles, with at least 100 mentions, reach the end of their lifespan around 24 hours after publication.}
  %  \label{fig:lifespan_hist}
%\end{figure}

%\subsection{Exploratory analysis}
%To investigate our dataset and validate some of our modelling assumptions, we perform a set of exploratory data analyses, which are shown in \ref{fig:lifespan_hist,fig:mention_dist}.
%The analysis of article publications time suggests that we should not use the time of the day as a feature for the prediction, because there could be artifacts from the data collection process. We also conclude that we should use at least one week of data for validation purposes because the publications on the weekend seem to differ from the remaining week.
In Fig. \ref{fig:lifespan_hist} we observe that the number of mentions on Twitter saturates generally within the first 24 hours after publication. We, therefore, define our prediction target as the number of mentions during 24 hours after publication.
We also found that there is a strong seasonality effect at weekends.
%In \ref{fig:mention_dist} we see that 95 percent of articles are mentioned less than 10 times within the first 29 hours after publication.

%\begin{figure}
%    \centering\small
%    \def\svgwidth{\columnwidth}
%    \input{time series.pdf_tex}\vspace*{0.5cm}
%    \caption{Distribution of the number of mentions per article over time during one week starting at 13.10.2018. The plot shows that most articles have a low number of mentions, which is a characteristic of average day in the cryptocurrency domain.}
%    \label{fig:mention_dist}
%\end{figure}

\section{Predictive analytics} \label{sec:problem_definition}

We aim to explore the different machine learning models on the task of predicting the total number of article mentions within a defined time horizon after the article's publication time.  
Let $t_0, t_1$ be two timestamps with $t_0 < t_1 < t_0 + \delta$, where we set $\delta$ to 24 hours in this paper. 
The task of our machine learning models is as follows: Given an article published at time $t_0$ and all tweets published between $t_0$ and $t_1$ that mention the article, predict the cumulative number of tweets mentioning the article between time $t_0$ and $t_0 + \delta$. Here $t_1$ is the prediction starting time, which represents how much historical data can be used to predict.

\subsection{Feature Extraction}

For our predictive models we use three sets of features, which we name \textit{time series}, \textit{content} and \textit{context} features.
\paragraph{Time series features} Let $t_0$ be the publication time of an article and $t_1$ the current time. Let $d$ a timestep size (set to 1 hour in our experiments). The time series feature $f_k$ is then given by the number of mentions of the article between $t_0$ and $t_0 + kd$ where $k \in \{ j \in \{1, 2, ...\} | t_0 + jd \leq t_1\}$. As an example suppose that $t_1$ is 3 hours after $t_0$ and the article is mentioned twice, once and three times in hours 1, 2 and 3 since publication respectively. Then there are 3 time series features: $f_1 = 2, f_2 = 3, f_3 = 6$.
Note that the number of these features is not constant, but depends on $t_1$ as defined in \ref{sec:problem_definition}.
\paragraph{Content features} We extract a vector of content features from each article, by using a keyword list to allow the models to learn individual dynamics for articles related to different cryptocurrencies. Each cryptocurrency related concept is represented by a binary feature, that is set to 1 if one of the keywords related to the concept is present in the title of the article. %The full list of keywords used can be found in Appendix A.
\paragraph{Context features} The amount of Twitter mentions might further be related to the publisher of an article. There seem to be very few publishers with significantly more mentions on Twitter than the other publishers. We count the total number of mentions of each publisher in our training set. We then extract the 10 publishers with the highest numbers of mentions. For each of these publishers, we introduce a binary feature set to 1 if the article was published by the respective publisher.

\subsection{Predictive Models}

\subsubsection{Baseline model}
As a baseline model, we use a linear extrapolation of the last $k$ time series features by fitting a linear function of the time step to the dataset given by $\{(K-k + 1, f_{K-k + 1}), ..., (K, f_{K})\}$. Here $K$ is the number of time series features available for the article. To predict the total number of mentions at the target time, we evaluate the model for the time step $N$, such that $Nd = \delta$. The model ignores content and context features. In our experiments, we will choose $k=3$.
%Note that this model does not take into account any previously observed Twitter dynamics. 

\subsubsection{Autoregressive models}
A common type of time series model is an autoregressive model \cite{timeseries}. An autoregressive model of order $k$ predicts the value at the next timestep $K + 1$ based on the values observed at the previous $k$ timesteps $K-k+1, ..., K$. In our experiments, we provide $K$ as an additional input to the model. The idea is, that the dynamics can be very different a few hours after the publication and shortly before the end of the prediction window. In some experiments, we will further provide content and/or context features to the model. In our case, we have to predict multiple steps in the future. This is achieved by first predicting a single timestep. We then assume that the predicted value is the correct value and use it as an input for the prediction of the next timestep. By recursively applying this strategy, we can predict an arbitrary amount of timesteps ahead.

For autoregressive models of order $k$, we generate multiple training samples from each time series $f_1, ..., f_N$. The first sample uses $f_1, ..., f_k$ to predict $f_{k +1}$. The second sample predicts $f_{k +2}$ from $f_2, ..., f_{k +1}$ and so on.

We use two different kinds of autoregressive models. The first one uses a linear model to predict the next timestep and the second one uses the random forest regressor.

\subsubsection{Random forest regressor}
% \paragraph{Linear model}
% Let $x_1, x_2, ..., x_n$ be a set of features and $y$ the target variable. The linear model assumes: $y = \beta_0 + \beta_1 x_1 + ... + \beta_n x_n + \epsilon$, where $\epsilon$ is an error term. Given a dataset, the parameters $\beta_0, ..., \beta_n$ can be estimated by assuming a Gaussian distribution of $\epsilon$ and then maximizing the likelihood. We use the implementation from the Python library scikit-learn.\footnote{\url{https://scikit-learn.org/stable/modules/generated/sklearn.linear_model.LinearRegression.html}}.
%
% \paragraph{Random forest regressor}
A random forest is an ensemble of decision trees. The total response of the random forest model is the average prediction of all decision trees. In order to increase the variety of the individual decision trees, each tree is trained on a bootstrapped sample from the original dataset and uses only random subsets of the features for each decision. For more details about random forests see \cite{random_forests} or the documentation of the scikit-learn implementation that we use for our experiments.\footnote{\url{https://scikit-learn.org/stable/modules/generated/sklearn.ensemble.RandomForestRegressor.html\#sklearn.ensemble.RandomForestRegressor}}

\subsubsection{Sequence-to-sequence model} As described in \cite{seq2seq}, sequence-to-sequence model consists of two recurrent neural networks (RNN). The first RNN is called the encoder. This recurrent network receives as inputs all available time series features. The outputs of this model are discarded. The second RNN is called the decoder. The initial hidden state is given by the final hidden state of the encoder. The first input to the decoder is the last time series feature. In our architecture, the output of the decoder at each timestep serves as input for a fully connected network with one hidden layer that outputs the predicted value for the next timestep. If context or content features are used, those features serve as additional inputs to the fully connected network. The predicted value is then used as the input at the next timestep. %The architecture is visualized in \ref{fig:seq2seq}. 
As a loss function, we use the sum of squared errors between the predicted values and the real values. Be $f_1, ..., f_l$ the input time series and $f_{l+1}, ..., f_K$ the real continuation. Be $\hat{f}_{K+1}, ...\hat{f}_N$ the predicted values. The loss for this time series is then given by $$ L = \sum_{i = l+1}^K (f_i - \hat{f_i})^2 $$

\paragraph{Gated recurrent units}
Our sequence-to-sequence model implementation is based on the gated recurrent unit (GRU), a variant of recurrent neural networks (RNN). The following definition of a GRU is taken from \cite{gru}. Let $\sigma(\cdot)$ denote the sigmoid function, $\tanh(\cdot)$ the hyperbolic tangent and $\odot$ the element-wise matrix multiplication operator. Be $x_1, x_2, ..., x_n$ a sequence of inputs (e.g. a time series) and $h_0$ the so-called \emph{initial hidden state} usually set to 0. The RNN using GRUs then outputs a sequence $h_1, h_2, ..., h_n$ defined by
\begin{align*}
    z_t &=  \sigma(W_z x_t + U_z h_{t-1})\\
    r_t &=  \sigma(W_r x_t + U_r h_{t-1})\\
    \tilde{h}_t &= \tanh(W x_t + U( r_t \odot h_{t-1})) \\
    h_t &=  (1 - z_t) \odot  h_{t-1} + z_t \odot \tilde{h}_t
\end{align*}
where $W_r, W_z, W, U_r, U_z, U$ are model parameters learned during training. We use the TensorFlow GRU implementation.\footnote{\url{https://www.tensorflow.org/api_docs/python/tf/nn/rnn_cell/GRUCell}}

%\begin{figure}
%    \centering
%    \def\svgwidth{\columnwidth}\scriptsize
%    \input{NN_Architecture_ink.pdf_tex}\vspace*{0.5cm}
%    \caption{The architecture of the sequence-to-sequence model. This model receives as input time series features $f_1, ..., f_l$ and outputs predicted time series values $\hat{f}_{l+1}, ..., \hat{f}_{K}$ Each box represents a neural network. The boxes labeled with 'E' are the cells of the encoder. The boxes labeled as 'D' belong to the decoder, while the boxes with 'Dense' label represent the fully connected layer that translates the decoder output into the next predicted value. The horizontal arrows between the encoder and decoder cells represent the propagation of the hidden state. The arrows entering the boxes from below represent the inputs to the neural networks. Arrows leaving the boxes towards the top represent outputs. Lines are dotted if repetitions of the same pattern are omitted.}
%    \label{fig:seq2seq}
%\end{figure}

\subsection{Evaluation set-up}

\subsubsection{Dataset}

In this paper, we have done the evaluation of our real-time model on 23535 articles published between 2018-12-02 00:00:00 and 2018-12-09 00:00:00 and all tweets mentioning those articles. The corresponding training set contains all articles published before 2018-12-02 00:00:00 with the related tweets. In total, the training set consists of 125248 articles from roughly 35 days.

\paragraph{Bootstrapping}
We want to estimate confidence intervals of the performance of the different models instead of just obtaining point estimates. Therefore, we use bootstrapping to generate 100 new datasets consisting of 2000 samples from the validation set.

\paragraph{Balancing} Because popular articles are rare, we balance the training set. Let $M = .05 * N_{train}$ where $N_{train}$ is the total number of articles in the training set. We first sort the training set in descending order by the number of mentions and then keep the top $M$ articles with most mentions. We also draw $M$ random samples from each of the sets of articles lying between the 75th and 95th percentile and below the 75th percentile. We hence obtain a training set with an equal number of high importance (above 95th percentile), medium importance (between 75th and 95th) and low importance (below 75th) articles.

\subsubsection{Models}
For the baseline model, we choose a linear interpolation of the most recent 3 time series features.
The linear autoregressive model is evaluated for orders 1, 3 and 5. 
The random forest autoregressive models are all of order 3 with 50 and 500 estimators.
The sequence-to-sequence model has a hidden state size of 200 for the encoder and the decoder. 
The hidden dense layer consists of 200 units. 
During training, we drop out 30 percent of the inputs to the dense layer as well as 10 percent of the hidden state passed to the next step of the RNN. 
The network is then trained in an end-to-end fashion, using back-propagation with training batches of size 64. 
The model is trained for 30 epochs with a learning rate .001, then for another 30 epochs with learning rate .0001 and finally for yet another 30 epochs using learning rate .000001 using Adam optimizer.

% As each training sample contains the whole time series data, a random prediction start time is chosen for each sample. 
% For example, one sample has to predict based on the data observed within the first 3 hours after publication, while in the same batch another sample uses the first 20 hours as the input for the prediction. 
% This is necessary, because the model later has to be able to start its prediction at any time point.

The baseline model is only provided the time series features. 
The autoregressive models and the sequence-to-sequence model are trained using the time series, content and context features.

\subsubsection{Evaluation metrics}\label{sec:evaluation}

The goal of our prediction is to extract the most relevant cryptocurrency related articles. Most articles accumulate very few mentions on Twitter. Those articles would heavily influence the performance scores because they represent the vast majority of articles in the validation set. However, the performance of those articles does not represent the performance with respect to our goal. For this reason, our evaluation focuses on the top k articles with most mentions on Twitter.

There are two different properties of the predictions that can be used to assess performance: the accuracy of the predicted number of mentions, and the quality of the induced ranking of articles. We use \textit{mean absolute percentage error} to measure the former property and \textit{normalized discounted cumulative gain} to measure the latter.
\paragraph{Mean absolute percentage error (MAPE)} The MAPE is used to evaluate the quality of the predicted value. It computes by how many percents the predicted value deviates from the actual value on average. The reason for using percentage errors instead of absolute errors lies in the great difference between numbers of mentions of articles. A metric based on absolute errors would most likely be dominated by the very few articles with significantly more mentions. The MAPE is defined as follows $$ \mathrm{MAPE} = \frac{1}{n} \sum_{i=1}^n \bigg|\frac{f_{\delta}^{(i)} - \hat{f}_{\delta}^{(i)}}{f_{\delta}^{(i)}}\bigg| $$
where $\delta=24$ denotes 24 hour window from the publication date.

\paragraph{Normalized discounted cumulative gain (NDCG)} The second interesting property of the predictions is the ordering of the articles induced by the predicted values. We could obtain the most important articles from a model that achieves poor performance with respect to the MAPE but yields a good approximation of the ordering of the news. The discounted cumulative gain (DCG) is high if the top k predicted articles achieve a high number of mentions. To compute the DCG, the articles are first ordered by their predicted importance such that $\hat{f}_{\delta}^{(1)}, \hat{f}_{\delta}^{(2)}, ..., \hat{f}_{\delta}^{(N)}$. Be  $f_{\delta}^{(1)}, f_{\delta}^{(2)}, ..., f_{\delta}^{(k)}$ the observed importance values of the first k articles from this ordered set. $\mathrm{DCG}_k$ is then defined as
$$
\mathrm{DCG}_k = \sum_{i=1}^k \frac{f_{\delta}^{(i)}}{\log_2(i + 1)}
$$
$\mathrm{IDCG}_k$ is now defined as the maximal achievable $\mathrm{DCG}_k$ which is computed as the $\mathrm{DCG}_k$ based on an ordering according to the observed instead of the predicted values. The $\mathrm{NDCG}$ can be computed as $$ \mathrm{NDCG}_k = \frac{\mathrm{DCG}_k}{\mathrm{IDCG}_k} $$
Hence the maximal achievable $\mathrm{NDCG}_k$ is 1.

\subsection{Results}
We will discuss the overall results of our models and compare their performance in the prediction of the number of mentions and the order of the articles. %Afterwards, we will have a look at some of the individual model and justify some of our choices for hyperparameters with experimental results.

% We compare the mean absolute percentage error (MAPE) and normalized discounted cumulative gain (NDCG), as defined in \ref{sec:evaluation}, for different prediction start times. 

We vary the prediction start time to be 5, 10, 15 or 20 hours after publication time while keeping the target prediction time fixed at $\delta=24$ hours after the publication. 
For instance, for a start time of 5 hours, the model gets five data points as input, describing the mentions in the first 5 hours. 
It then predicts the number of mentions after another 19 hours. 
Similarly, for a start-time of 15 hours, the model gets 15 points as input and has to predict 9 hours into the future.

% Of course, predictions get better the closer the prediction start time gets to the target prediction time, because the later the start time the more data already has been observed and the less steps in the future have to be predicted. 
% However, we are interested how the models compare at different point in the lifetime of an article.

\begin{figure}[h]
  \centering
  \includegraphics[width=0.95\linewidth]{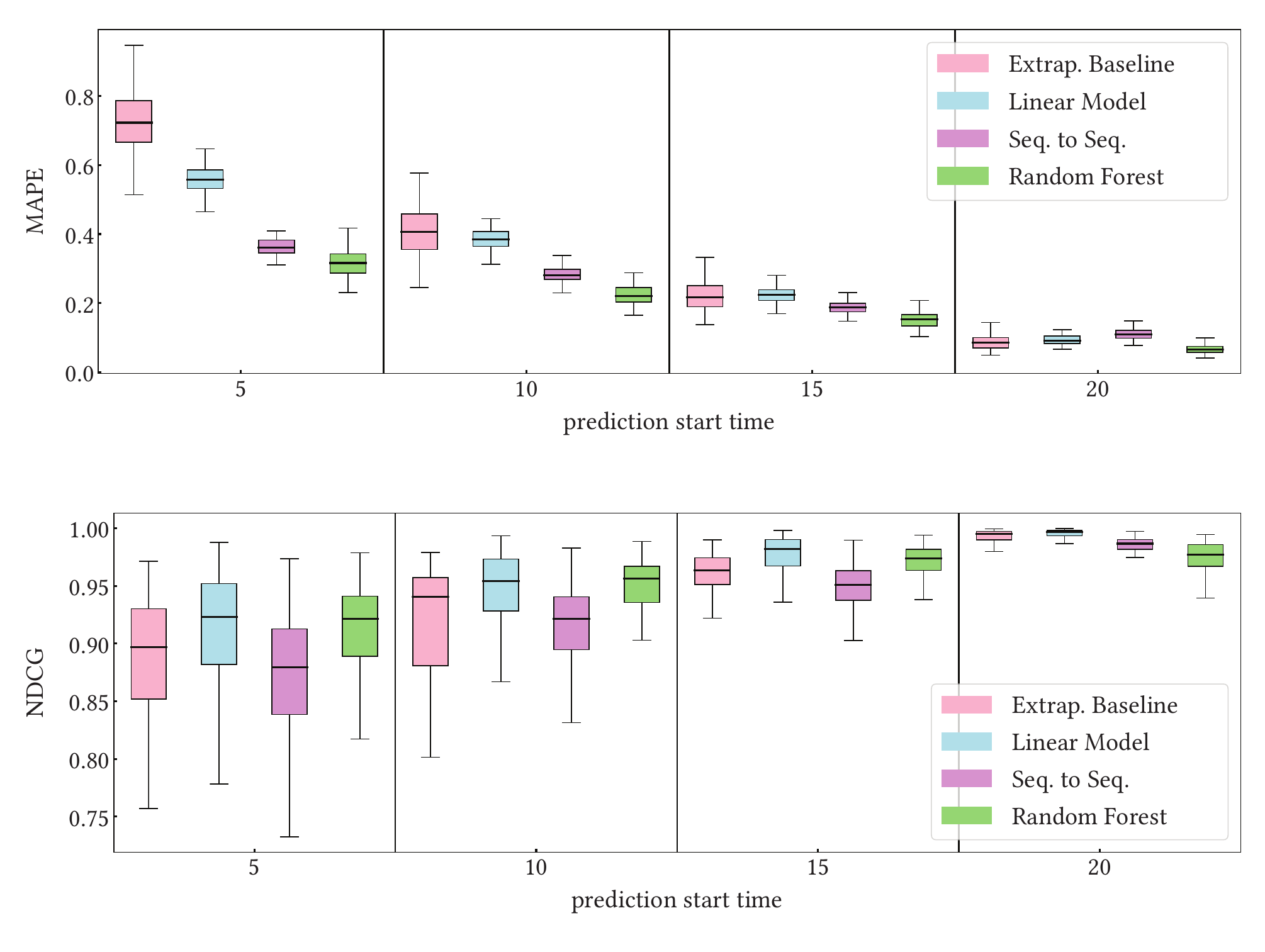}
 \caption{MAPE and NDCG of the different models evaluated on the test data set. The quantiles are determined using 100 bootstrap samples. Predictions are evaluated at 4 different prediction times, 5, 10, 15, and 20 hours after publication of an article.}
    \label{fig:model_comparison}
\end{figure}

\subsubsection{Overall performance}

In Fig. \ref{fig:model_comparison} we show the experimental results of the baseline model, autoregressive model, random forest autoregressive model and sequence-to-sequence model on the test dataset.

As expected, we see that all models make better predictions, the closer the prediction start time is to the target time. 
After 15 and 20 hours from the publication time, the baseline model already achieves very good performance with MAPE of $20\%$ and less. 
As we have seen, the number of mentions in most news articles starts to saturate after $10-20$ hours. 
Because of this, the linear extrapolation that is performed by the baseline model can be quite accurate at later prediction points. 
At prediction start times of 15 and 20 hours,
% RF and S2S models are overall comparable to the baseline. 
the random forest (RF) and sequence-to-sequence (S2S) model achieve a slightly lower MAPE than the baseline.

However, we are more interested in the early prediction start points. Ideally, we want to make an accurate prediction about the popularity of an article as soon as possible after its publication. 
At prediction start time 5 hours after publication advanced models achieve a significantly lower MAPE than the baseline. 
RF and S2S model achieve a MAPE around $30-40\%$, while the linear model is at about $45\%$ and the baseline at $70\%$. For predictions starting 10 hours after publication, the baseline and the linear model improve significantly over their performance 5 hours after publication. 
However, RF and S2S model still achieve a significantly lower MAPE.

Overall, we can say that RF and S2S model is able to achieve significantly lower error rates close to the publication time of an article. 
All models are comparable about 20 hours after publication. 
The RF model achieves the lowest MAPE overall.

\begin{figure}[h]
  \includegraphics[width=0.95\linewidth]{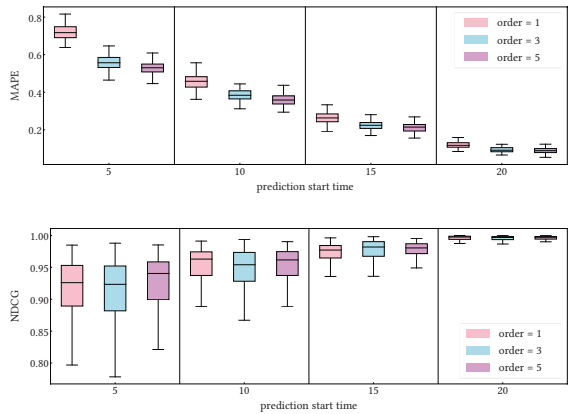}
    \caption{MAPE and NDCG of a linear model of orders 1, 3 and 5 evaluated on the test dataset. The quantiles are determined using 100 bootstrap samples. Predictions are evaluated at 4 different prediction times, 5, 10, 15, and 20 hours after publication of an article.}
    \label{fig:order_comp}
\end{figure}

\begin{figure}[h]
  \includegraphics[width=0.95\linewidth]{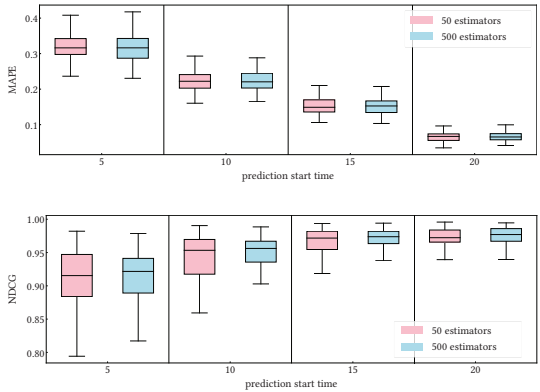}
   \caption{MAPE and NDCG of a random forest model with 50 and 500 estimators evaluated on the test dataset. The quantiles are determined using 100 bootstrap samples. Predictions are evaluated at 4 different prediction times, 5, 10, 15, and 20 hours after publication of an article.}
    \label{fig:n_estimators_comp}
\end{figure}

\begin{figure}[h]
  \includegraphics[width=0.95\linewidth]{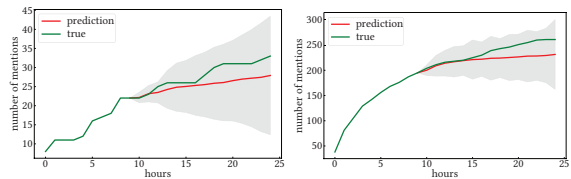}
 \caption{Predictions of the RFAR model for two articles from the test set. The shaded area shows $95\%$ prediction intervals that are determined from the distribution of the predictions by the ensemble estimators.}
    \label{fig:prediction_intervals}
\end{figure}

It is instructive to look at the NDCG as well. 
Here we do not see any model being significantly better than the baseline model, which is mainly due to the fact that the baseline model is already very good at predicting the final order of the news articles shortly after publication. 
It achieves an NDCG of around 0.9 only 5 hours after publication. The S2S model seems to perform worse than the other two models in predicting the order of the articles. 
This is likely due to the fact, that it has a large degree of freedom in the model parameter space and probably the rank regularization could help. 
However, additional tuning of the objective function of the S2S model was left for future work.

%This is likely due to the fact, that we did not properly tune the S2S model parameters. 
%The RF is comparable to the linear model, but it seems to be significantly worse 20 hours after publication.

Predicting a rough order of popular news articles seems to be possible by just linearly extrapolating the number of mentions in the first few hours. 
Improving over that is hard due to the high uncertainty of the predictions of the number of mentions after 24 hours. 
Looking at the performance of the linear model and the RF, it seems to be feasible to improve over the baseline, but the uncertainty of the predictions remains high.

% Based on the results discussed in this section, we choose the random forest model to implement in our online-prediction system (see \ref{sec:online}).

\subsubsection{Effect of order}

For all autoregressive models, we have to choose an order. A higher order increases computational cost but potentially also the prediction quality. 
To measure the effect on the prediction quality, we evaluate the linear autoregressive model on different orders 1, 3 and 5.

The results are shown in Fig. \ref{fig:order_comp}. 
We find that for each prediction start time the MAPE of the order = $3$ model is significantly lower than the MAPE of the order = $1$ model. 
However, increasing the order of the model to $5$ does not seem to give a significant error reduction. 
The NDCG is not significantly different between the different choices of the order.

In view of these results, we choose an order of $3$ for all autoregressive models from here on. 
While we might be able to gain slightly better predictions by choosing an order of more than $5$, choosing $3$ seems to be a good compromise between model performance and computational cost.

%\subsubsection{Random forest autoregressive model}

Random forest models provide a number of hyper-parameters such as the number of estimators, the depth of the tree or the size of the leaf nodes.
To this end, we compare a random forest autoregressor with 50 to one with 500 estimators. The results, depicted in Fig. \ref{fig:n_estimators_comp}, show no significant different in MAPE or NDCG for both models. 
Choosing 500 instead of 50 estimators somewhat decreases the variance of the NDCG. 
For the random forest models in our other experiments, we will, therefore, use 500 estimators, which are still manageable in computation. 
Experimenting with other hyperparameters was out of the scope of this work and is left for future work. 

%While it was out of scope to tune all of these, we do want to investigate the number of estimators, which usually has the largest influence on the performance of the random forest.
%\subsubsection{Sequence-to-sequence model}
%It was out of scope of this project to extensively tune the hyperparamters of the neural network. We set the hyperparameters based on some preliminary experiments, but it is likely that proper hyperparameter tuning would lead to better performance of the sequence to sequence model.

\subsubsection{Uncertainty prediction}

In addition to achieving the best model performance in our experiments, the random forest model also gives us a natural way to quantify the prediction uncertainty. 
Instead of just calculating the mean of the ensemble predictions, we can calculate percentiles of the predictions to get prediction intervals with $95\%$ coverage. 
This is shown in two example time series in Fig. \ref{fig:prediction_intervals}.

\subsection{Model Deployment}\label{sec:online}

We use the trained autoregressive model to do online predictions on real-time data. 
The data extraction server\footnote{deployed on the Google Cloud (GC)} constantly retrieves new tweets and articles and finds the tweet-article matchings, which are saved as a new batch into an online database\footnote{MongoDB deployed on Amazon Web Services (AWS)}. 
We generate the dataset for prediction by querying the database for articles published in the last 24 hours. 
The queried data is then pre-processed in order to extract time series, content, and context features, as described earlier.

Then, we predict the importance values of new articles at 24 hours after the article's publication time, using the previously trained model. 
The prediction is performed every 10 minutes.
% and is scheduled with the Advanced Python Schedule package.\footnote{\url{https://apscheduler.readthedocs.io/en/latest/}} Moreover, at each prediction update, the predicted importance values are saved in a JSON file data. 

The current predicted importance values are visualized in an interactive webpage\footnote{Link to webpage: \url{http://cryptodatathon.com/ranknews}}. 
The web application is developed using Python-based Flask web development framework\footnote{\url{http://flask.pocoo.org/}}: the front-end makes constantly Ajax GET requests to the back-end, which reads the updated importance values from JSON files. 
For data visualization, we generate interactive charts with the JavaScript C3.js library\footnote{\url{https://c3js.org}}. % Finally, the application is deployed on an AWS server with Apache and WSGI.
The web front-end provides threes tabs (i-iii), showing the following dynamic content.
(i)  A table with the most important articles can be sorted by the current number of tweets mentions or by the predicted importance value at the target of 24 hours after publication. 
It also embeds the link to each article and displays some relevant information about the article, such as the publisher and publication time.
(ii) A combined line chart shows the cumulative time series of the most important articles - a continuous line for the observed hourly mentions and a dashed line for the predicted values up to 24 hours since the article's publication (see Fig.\ref{fig:webChart}).
(iii) Two pie charts present the statistics related to article publisher (context) and cryptocurrency (content) based keywords (see Fig.\ref{fig:webStats}).\\

\section{Conclusion}
In this paper, we introduce an online data mining system relating cryptocurrency news to the tweets discussing them.
This data pipeline paves the way for monitoring cryptocurrency news of public's interest, identifying and predicting poplar news, and tracking public opinions towards cryptocurrencies.   

Data exploration on the collected paired news articles and tweets characterized top publishers, top cryptocurrencies discussed on Twitter as well as the lifespan of these news discussions.
We also perform preliminary predictive analytics using machine learning and deep learning models.
This work is a first step towards providing a prediction system, that detects articles that are going to become popular shortly after they are published. 

Our current system still needs to observe a few hours of data before making a prediction.
For future work, the goal would be to make more accurate predictions within the first hour after publication. 
This is possible, by exploring different representations of the article content by more advanced NLP models.

% We analyse predictive ability of first 24 hours after the article publication time. Finally, we used the predictions to create a real-time system for monitoring and tracking articles and their corresponding future mentions.

%We found that machine learning models achieve significantly higher prediction accuracy than a naive baseline. However, the quality of predicting the order of popular news is only slightly above the baseline, in our prediction setting.

%Furthermore, we plan to open-source the gathered and processed data for community in real-time.

%Furthermore, our current online-prediction model is trained offline and only makes predictions in real-time. In the future, we want the model to be able to learn online from the data it observes. Additionally, we want the model to be evaluated in real-time, giving the user information about how the model performs over time and even after its deployment.

%Finding relevant information in the overwhelming amount of news articles that are published every day is an important task not just restricted to cryptocurrencies. If the goal of accurately predicting news popularity early on proves to be achievable, this will open up a wide range of applications in other domains as well. Ultimately, we hope that such a system can help people to make better decisions in today's complex world.

%\FloatBarrier

\section{Acknowledgments}
N.A.-F. and T.G. are grateful for financial support from the
EU Horizon 2020 projects: SoBigData under grant agreement
No. 654024. J.B., R.H. and D.L are grateful for the support of professor A. Krause on the Data Science Lab 2018 course at ETH Zurich.

\end{document}